%% file: paper.tex
\documentclass[runningheads]{llncs}

\usepackage[utf8]{inputenc}
\usepackage[T1]{fontenc}
\usepackage{lmodern} 
\usepackage{makeidx}  
\usepackage{subfig}
\usepackage{graphicx}
\usepackage[ruled,linesnumbered]{algorithm2e}

\usepackage{etex}
\usepackage[hidelinks]{hyperref}

\usepackage{bm}
\usepackage{amsmath,amssymb}
\usepackage{mathabx}
\usepackage{wasysym}
\usepackage{booktabs}
\usepackage[normalem]{ulem}
\usepackage{colortbl}
\usepackage{listings,xcolor,multicol}
\usepackage{chngcntr}
\usepackage{url}
\usepackage[defblank]{paralist}
\usepackage{fancybox}
\usepackage{float}
\usepackage{xspace}
\usepackage{stmaryrd}
\usepackage{tikz}
\usetikzlibrary{positioning,shapes,snakes,fit,patterns,chains,automata,calc}

\newcommand{\sage}{\textsc{SaGe}}
\newcommand{\sagee}{\textsc{SaGe}\xspace}

\hypersetup{
  pdftitle={SaGe: Preemptive Query Execution for High Data Availability on the Web},
  pdfsubject={SaGe: Preemptive Query Execution for High Data Availability on the Web},
  pdfauthor={Thomas Minier, Hala Skaf-Molli, Pascal Molli},
  pdfkeywords={Semantic Web, SPARQL query processing, Data availability, Preemptive query execution}
}

\newcommand*{\medcup}{\mathbin{\scalebox{0.7}{\ensuremath{\bigcup}}}}

\begin{document}

\title{\sage: Preemptive Query Execution for High Data Availability on the Web}
\titlerunning{\sage: Preemptive Query Execution for High Data Availability on the Web}  
%
\author{Thomas Minier \and Hala Skaf-Molli \and Pascal Molli}
\authorrunning{Thomas Minier et al.} 
%
\tocauthor{Thomas Minier, Hala Skaf-Molli, Pascal Molli}
\institute{LS2N, University of Nantes,
Nantes, France\\
\email{firstname.lastname@univ-nantes.fr}
}

\maketitle              

\begin{abstract}
  Semantic Web applications require querying available RDF Data  with high performance and reliability.
  However, ensuring both data availability and performant SPARQL query execution in the context of   public
  SPARQL servers  are challenging problems. Queries could have arbitrary execution time
  and unknown arrival rates. In this paper, we propose \sagee, a preemptive server-side SPARQL query engine.
 \sagee relies on a preemptable physical query execution plan and preemptable physical operators.
 \sagee stops query execution after a given slice of time,
  saves the state of the plan and sends the saved  plan back to the client with retrieved results.
  Later, the client can continue the query execution by
  resubmitting the saved  plan to the server. By ensuring  a fair query execution,
  \sagee maintains server availability and provides high query throughput.
  Experimental results demonstrate that \sagee
  outperforms the state of the art  SPARQL query engines in terms of query throughput,
  query timeout and answer completeness.
\end{abstract}

\section{Introduction}\label{sec:introduction}

The semantic web is a global unbound data space where data providers
publish data in RDF and data consumers execute SPARQL query though
semantic web applications~\cite{bizer2009linked}. When writing a
semantic web application, it is crucial that RDF data are
available and SPARQL queries execution are performant and
reliable. However, ensuring both RDF data availability and query
performance is a major issue for the semantic web.

A semantic web application can rely on public SPARQL endpoints to
access RDF data. However, as reported in \cite{buil2013sparql}, during
27 month monitoring, only 32.2\% of public endpoints have a monthly
"two-nines'' up-times.  Undoubtedly, this is a problem for writing semantic
web applications. This is also a problem for data providers that have
to support an unpredictable load of arbitrary SPARQL queries. Public
SPARQL endpoints protect themselves by using quotas in time and query
results as in DBPedia SPARQL
endpoint~\footnote{\url{http://wiki.dbpedia.org/public-sparql-endpoint}}. Such
protections drastically limit the availability of RDF data when executing long-running queries
pushing developers to copy data locally and query data dumps.

The Public SPARQL endpoints are not the only way to query the semantic
Web. Various tradeoffs have been explored with Link
Traversal~\cite{olaf2009} or Linked Data Fragments
(LDF)~\cite{verborgh2016triple} as reported in~\cite{olaf2017}. The LDF approach
demonstrates how SPARQL query execution can be distributed between
data providers and data consumers to improve data availability. The interface of the public LDF
servers can scale at low cost for data
providers because this interface only processes constant time operations, such as paginated triple
patterns with the TPF interface~\cite{verborgh2016triple}.
However, costly SPARQL operations, like joins, are performed on client side.
As a large number of  intermediate results are transferred to the client, the performances of the query execution can be significantly degraded
compared to the performances of public SPARQL endpoints. Consequently,
writing web applications with low performances SPARQL query execution
remains a serious limitation for the development of the semantic web.
The main challenge is to find \emph{an interface for public servers and a query
execution model ensuring both RDF data availability at low cost for data
providers and high query execution performances for semantic web application
developpers}.

In this paper, we propose \sage, a new SPARQL query engine based on
stateless preemptable query plans. The main idea is to allow a \sagee
public server to preempt a query execution after a predefined slice of
time, save the query execution state, and send this state to the
smart \sagee client. Later, the client is free to continue execution by
resubmitting the saved execution state. This execution model allows
complex queries to be executed without explicit and costly pagination
performed by clients, based on \emph{Limit/Offset/OrderBy}  query rewriting techniques~\cite{axel2014}.
The time quota allows queries with different number of results and different execution time to run on the same server while ensuring \emph{proportional fairness}
and a \emph{starvation-free} for queries~\cite{silberschatz2014operating}.
Finally, as query execution states are stored in the \sagee client, queries can be resumed even
after a failure or a timeout from the \sagee server.
The contributions of this paper are:
\begin{asparaenum}

\item We outline practical limitations of  public SPARQL processing models: SPARQL endpoints and TPF servers.
Availability and performance issues prevent the usage of existing infrastructures in real-world semantic web applications.

\item We propose \sagee, a \emph{stateless preemptable query engine} that combines both proportional fairness of
TPF and performances of SPARQL endpoints.

\item We formalize preemptable query execution plan and present a set of physical operators that allow a preemptable
  execution of BGP. We present the \sagee implementation for preemptable query
  execution plan and preemptable iterators for join processing.

\item We evaluate \sagee by running extensive experimentations using \emph{WatDiv}~\cite{alucc2014diversified}.
    Results suggest that \sagee query engine improves query throughput and
    query timeout compared to SPARQL endpoints and TPF approaches.
\end{asparaenum}

This paper is organized as follows.  Section~\ref{sec:related_work}
summarizes related works.  Section~\ref{sec:approach} presents the
\sagee query execution model and the formalization of the preemptable
query execution plan and physical operators. Section~\ref{sec:engine}
details \sagee query optimizer and query engine.
Section~\ref{sec:exp_study} presents our experimental results.
Finally, conclusions and future work are outlined in
Section~\ref{sec:conclusion}.

\section{Related Work}\label{sec:related_work}


\noindent \emph{Public SPARQL endpoints} allow data consumers and semantic web
applications to execute \emph{expressive} SPARQL
queries without copying data locally. However, as
these endpoints are exposed to an unpredictable load of arbitrary
SPARQL queries, they enforce a fair use policy of
server resources by relying on \emph{server quotas}.
These quotas restrict the time for executing a query in the server,
the maximum number of results per query or the rate at which clients send queries.
 The DBpedia public SPARQL endpoint restricts SPARQL query
execution time to 120 seconds, the maximum number of results to 2000 and the estimated cost of queries
to 1500 seconds~\footnote{\url{http://wiki.dbpedia.org/public-sparql-endpoint}}.
The rate of queries  limits the number of queries that a single IP can send to the server
during a period of time. This is a common technique to protect public HTTP servers against DOS attacks.
Although, these quotas are required to provide stable and responsive
endpoints for the community, the execution of complex queries under
these quotas is more challenging for data consumers and semantic web applications. If a query  execution exceed these quotas,
 the query  has to be paginated using \emph{Limit/Offset/OrderBy}
rewriting techniques~\cite{axel2014}.
However, these techniques may require fine tuning and could deteriorate performance.




\noindent \emph{The Triple Pattern Fragments (TPF)~\cite{verborgh2016triple}}
proposes an alternative approach to consume Linked Data by distributing SPARQL
query processing between clients and servers. The TPF server
evaluates only triple patterns and retrieves paginated results.
To execute a full SPARL query, a TPF client decomposes it
into a sequence of triple patterns queries, send them to the TPF
server, collects pages of results and performs all others operations,
like joins, locally. By processing only triple
patterns in near constant time~\cite{fernandez2013binary},
TPF servers fairly allocate resources to their clients  without the need of
server quotas, but they still need to limit the rate at which clients access the Web server.
Unlike SPARQL endpoints, a TPF server has a simple interface that does not differentiate
between simple and complex queries.
Developers do not need to paginate queries themselves
to bypass server quotas, as it is the server that handles pagination.
The downside is that complex queries requires much more HTTP requests to
the TPF server than the simple ones, which is a form of proportional
fairness~\cite{WIERMAN201139}.
As joins are performed on client side, all intermediate results are
transferred to client side. Consequently, the overall data transfer
from TPF servers to TPF clients leads to poor query processing
performance~\cite{olaf2017}.
Different LDF server
interfaces~\cite{hartig2016bindings,van2015substring,vander2015opportunistic}
have been proposed to reduce the number of subqueries required to
evaluate SPARQL queries by increasing the expressivity of
the TPF server interface. However, as some SPARQL operations are still
executed client-side, query performance is still deteriorated by the transfer of intermediate results.



\noindent \emph{Preemption} is a general approach to provide a fair use policy of
resources. It is commonly used in operating system, network and
databases, and heavily studied in both queueing theory~\cite{brockmeyer1948life} and
scheduling~\cite{silberschatz2014operating}. Considering one processor, a FIFO
queue of waiting tasks and a task arrival rate, a basic preemptive
scheduler stops the running task after a slice of time, saves the
state of task in the waiting queue and runs the next task.
This technique is called \emph{round robin scheduling} and ensures that
the system is \emph{starvation free}, \emph{i.e.},
a long-running task cannot block a short one in the queue.
Consequently, the throughput of the system is increased compared to
a system with no preemption, \emph{i.e.}, with an infinite time quota.
Preemptive query execution has been studied in the context of database
management systems (DBMS)~\cite{Stonebraker81}, where the DBMS support
multitasking to increase query throughput. A TPF server does not  need to support
preemption mechanism because it returns  one page of results for a
triple pattern is nearly in constant time~\cite{fernandez2013binary}.
Consequently, each task in the server's waiting queue has nearly the same execution time.
SPARQL endpoints support multitasking \emph{e.g.}, a Virtuoso server can
run several queries in parallel in different threads using preemption.
However, such preemption is only used for running queries, and not for the queries in the waiting queue.
In this case, the server can quickly be congested with long-running queries,
as they occupy the server threads, deteriorating the query throughput.


\noindent By analyzing existing public SPARQL processing approaches, we admit
the poor availability of RDF data for complex query processing.  On
one the hand, public SPARQL endpoints rely on server-side quotas to
diminish the impact of complex queries on the server performance,
reducing consequently the number of queries that can get complete results.
On the other hand, the TPF approach does not require such server-side quotas
and can process any query. However, complex queries generate a large
number of intermediate results that degrades drastically query execution performance.
Consequently, the limitations of existing approaches push
semantic web application developers to make a local copy of RDF data and execute queries locally.
This paper proposes a solution for public SPARQL processing for bypassing the problem of congestion of
public servers with complex queries. Preemption is a good solution for fair resources sharing
for unpredictable load or arbitrary queries.



\section{\sagee Approach}\label{sec:approach}

\sagee is a \emph{stateless preemptable SPARQL query execution} based
on \emph{time sharing} principles. This new execution model combines
both proportional fairness of TPF and performances of SPARQL
endpoints. A \sagee server executes a Basic Graph Pattern query for a
fixed slice of time, called \emph{time quota}, and returns a page of
results, with variable size, combined with the state of the resumable
physical plan of the BGP. Compared to the TPF approach, the
\emph{execution of a BGP} instead of triple patterns queries reduces
the number of intermediate results. The \emph{quota of time and the
  variable size pagination} discharge developers of the burden of
query rewriting. Finally, the \emph{stateless preemption} prevents the
problem of congestion of complex queries in the server.

During this work, we made the following hypotheses:
\begin{inparaenum}[(i)]
  \item \sagee servers are single writers, \emph{i.e.},
  all updates are controlled by the data providers through revisions.
  Consequently, queries are executed on immutable versioned datasets.
\item Versioned datasets are efficiently indexed by the data
  providers. In this paper, we rely on
  HDT~\cite{fernandez2013binary} for indexing and storing RDF
  data.
  \item In this paper, we focus on the execution Basic Graph Patterns (BGP queries), \emph{i.e.},
  conjunctive queries of triple patterns.
\end{inparaenum}

\subsection{\sagee Query execution Model}

\begin{figure}[t]
  \centering
  \resizebox{0.75\textwidth}{!}{%
    \input{./figures/sage_example.tex}
  }
  \caption{The \sagee query execution model}
  \label{fig:sage_model}
\end{figure}
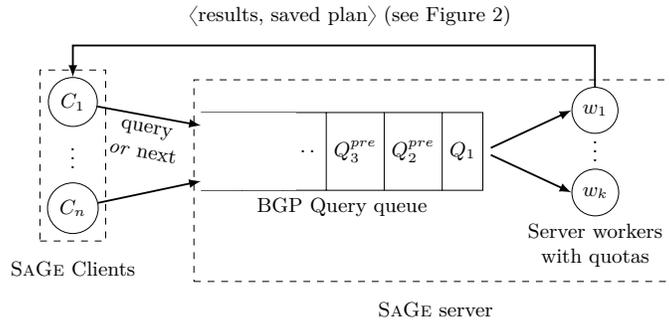

Figure~\ref{fig:sage_model} illustrates the \sagee model. As for TPF,
\sagee relies on a \sagee smart client and a light \sagee server. The
architecture of the  server follows the same architecture as Web
servers: a pool of \emph{server workers} are in charge of query
execution, and a \emph{query queue} is used to store incoming BGP
queries when all workers are busy. The queue contains one new incoming
query $Q_1$, \emph{i.e.}, a query that has not been executed by the server previously,
and two preempted queries $Q_2^{pre}$ and $Q_3^{pre}$, \emph{i.e.}, queries with
interrupted physical plans.

The \sagee client starts by submitting a BGP query $Q$ to the \sagee
server.  The query is added to the queue until a server worker is
available to process it.  When a worker is available, the server
computes a \emph{preemptable physical query execution plan} for $Q$.
A preemptable physical plan allows to evaluate $Q$ while supporting
preemption.  The \sagee server executes the preemptable physical plan
as follows:

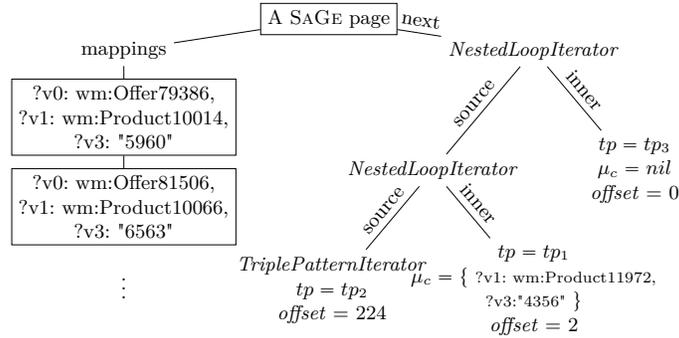
\begin{figure}[t]
  \centering
  \subfloat [BGP query $Q_1$, extracted from the WatDiv benchmark~\cite{alucc2014diversified}] {\label{query:q1}
\begin{lstlisting}[
       basicstyle=\scriptsize\sffamily,
       language=sparql,
       extendedchars=true
    ]
PREFIX wm: <http://db.uwaterloo.ca/~galuc/wsdbm/>
SELECT DISTINCT * WHERE {
 ?v0 <http://purl.org/goodrelations/includes> ?v1. # tp1
 ?v1 <http://schema.org/contentSize/contentSize> ?v3. # tp2
 ?v0 <http://schema.org/contentSize/eligibleRegion> wm:Country9. # tp3
}
\end{lstlisting}
  }\vfill
  \subfloat [One page returned by the \sagee server during $Q_1$ evaluation] {\label{fig:sage_page}
    \resizebox{0.75\textwidth}{!}{%
      \input{./figures/q1_page.tex}
    }
  }
  \caption{A tree representation of a page returned by the \sagee server when executing $Q_1$
    on the WatDiv Dataset}
  \label{fig:q1_page}
\end{figure}

\begin{asparaenum}

\item The server executes the plan until the quota is
  exhausted. Next, the \sagee query engine interrupts the execution
  and saves the current state of the query execution plan. This
  corresponds to the Round-Robin Scheduling algorithm that is
  starvation-free and preserves
  fairness~\cite{silberschatz2014operating}[section 6.3.4]. Others
  scheduling algorithms are also adequate, we chosen Round-Robin for
  its simplicity.

\item The server builds a \emph{page of results} using all retrieved
  results solution mappings, and a hypermedia link \texttt{next},
  which contains the saved state of the query execution plan. This
  page is described in the Figure~\ref{fig:q1_page} (we explain how
  this saved state is build in Section~\ref{sec:sage_plan}).  Then,
  the page is returned to the client. The \sagee server is fully
  stateless, \emph{i.e.} the saved state of the plan is not saved by
  the server.

\item When the results are received, the \sagee client is able to
  continue $Q$ execution with a fresh quota by submitting the saved
  execution plan back to the server, using the \texttt{next} link,
  which resume the preemptive execution of $Q$. If no \texttt{next}
  link is received, then the client knows that $Q$ has been fully
  executed, \emph{i.e.}, the client had received the complete results
  of $Q$.
\end{asparaenum}

\noindent The \sagee execution model has several advantages:

\begin{asparadesc}
\item[BGP on server side:] Compared to TPF, the BGP are processed on
  server side, the intermediate results are no more transmitted to the
  client, so the \sagee client doest not compute join operators, but
  just follow the $next$ link to complete the query execution.

\item[Constant arrival rate:] While executing a query, a client
  submits only one HTTP request at time to the server. 
 The client  gets the results for the query sequentially, by following
  the \texttt{next} links provided by the server, as with a classic
  REST collection. Thus, at a given time, a client cannot have more
  than one request in the waiting queue of the \sagee server. This is
  not the case with TPF, where a client can have several pending
  queries in the TPF server.


\item[Proportional fairness:] Thanks to preemption and constant
  arrival rate, \sagee executes BGPs with \emph{proportional
    fairness}, which increase \emph{query throughput} under high load.
  Proportional fairness means ‘fair’ for the response time of queries
  to be proportional to the queries complexity~\cite{WIERMAN201139},
  \emph{i.e.}, evaluation of long-running queries just require more
  calls to the \sagee server than short queries.

\item[Stateless Server:] Finally, the \sagee server is fully
  stateless, \emph{i.e.}, stopped query execution plans are sent back
  to the clients. Consequently, a long-running query cannot stay in
  the waiting queue. It exits the server to re-enter the waiting queue
  again, when the client uses the \texttt{next} link.  This allows a
  fair access to the waiting queue of the \sagee server and releases
  the server from storage overhead.
%
  Moreover, saving the state of query execution plans client-side
  allows the \sagee client to tolerate server failures.  If a call to
  the server is failed, the client can retry later. This opens the
  opportunities for client-side
  load-balancing~\cite{minierESWC2018ulysses}.
\end{asparadesc}

\subsection{\sagee Requirements}
\label{sec:rec}

To be effective and avoid performance deterioration during query execution, the preemptive execution
performed by \sagee
requires a fair value for the time quota and low overhead in time and space complexity for the preemption.

\begin{inparadesc}
\item[Fair value for the time quota] Finding the fair value for the
  time quota is important for performance. This value depends on query workloads.
  If the time quota is extremely small, the preemption
   impacts negatively all queries. Queries require more
  HTTP requests to be completely evaluated. In contrast, if the time
  quota is extremely large, the time sharing approach
  will degenerate to a FIFO policy and the server throughput is deteriorated
  \emph{i.e.}, long-running queries will impact negatively short ones. According to~\cite{silberschatz2014operating}[section
  6.3.4], \emph{a rule of thumb is that 80 percent of the CPU bursts
    should be shorter than the time quantum}, and the preemption overhead is a small fraction of the time quantum.
    For \sagee, in the experimental study (see Section~\ref{sec:exp_study}), we computed the time quota such that 80\% of queries of workloads
  are executed under the time quota,  the time quota is around $75ms$.
  We check the accuracy of the $80\%$ rule in the workload
  and ensure that the overhead of preemption (saving and loading the
  preempted plan) represent less than $10\%$ of the time quota.

\item[Low Preemption overhead] As the state of the physical plan is
  sent to the client, its space complexity must be bound by the
  complexity of the BGP query $O(|BGP|)$, \emph{i.e.}, the number of
  triple patterns in the BGP query.  Moreover, the time complexity for
  stopping, saving and reloading a preemptable physical execution plan
  should be negligible compared to the time quota itself.  This
  complexity must also be in $O(|BGP|)$.
\end{inparadesc}

\subsection{Formalization of Preemptable Physical Query Execution Plan}\label{sec:preempt_plan}

The preemptable physical query execution  and the corresponding join operators
used in \sagee support three
functions: \texttt{Stop}, \texttt{Save} and \texttt{Load}.
Definition~\ref{def:preemptable_plan} and Definition~\ref{def:preemptable_operator}
gives the specifications and the properties of these three functions
for preemtable physical query execution plan and its join operators, respectively.

\begin{definition}[Preemptable physical query execution plan]\label{def:preemptable_plan}

  Given a BGP $B = \{ tp_1, \dots, tp_m \}$ and a RDF dataset
  $\mathcal{D}$.  A preemptable physical query execution plan for $B$
  is a physical query execution plan that allows to evaluate $B$ over
  $\mathcal{D}$ and is composed by join preemptable physical
  operators. The plan supports the following functions:
  \begin{asparaitem}
  \item \texttt{Stop}: interrupts the plan in a correct state,
    \emph{i.e.} waits for all physical operators to have finished their critical sections.
    This function is executed in $O(|B|)$ time complexity.
  \item \texttt{Save}: serializes the correct state of the plan
    obtained by the \texttt{Stop} function. The space complexity of the
    serialized state is in $O(|B|)$.
  \item \texttt{Load}: reloads the plan in a correct state using the serialized state
    obtained by the last \texttt{Save} function.
    This function is in $O(|B|)$ time complexity.
  \end{asparaitem}
\end{definition}

The time and space complexity of \texttt{Stop}, \texttt{Save} and \texttt{Load} functions  determine the overhead of the preemption by the query engine.
Intuitively, this overhead must be negligible compared to the time quota allocated for query execution,
to avoid deterioration of query performance.

\begin{definition}[Preemptable physical join operators]\label{def:preemptable_operator}
  A preemptable physical join operator is a physical query operator
  that performs join processing and supports the following functions:
  \begin{asparaitem}
  \item \texttt{Stop}: interrupts the join operator in a correct state,
    \emph{i.e.} waits the physical operator to have completed its critical
    section. This function executes in $O(1)$ time complexity.
  \item \texttt{Save}: serializes the correct state of the operator
    obtained by the \texttt{Stop} function. The space complexity of the
    serialization is $O(1)$.
  \item \texttt{Load}: reloads the join operator in a correct state
    using a serialized state obtained by the last
    \texttt{Save} function.
    This function is in $O(1)$ time complexity.
  \end{asparaitem}
\end{definition}

According to Definition~\ref{def:preemptable_operator},
not all possible join operators can be implemented as preemptable join operators.
For example, hash joins based operators must maintain an internal state
which size depends on data complexity~\cite{PangCL93}, \emph{i.e.}, the number of triples in the RDF dataset.
Thus, the correct state of such operator cannot be serialized in $O(1)$.

These restrictions limit the choices of join operators for the \sagee query optimizer,
as it can only use preemptable join operators in the preemptable physical query execution plan.
Thus, the optimizer is limited in term of the \emph{shapes} of the plans it can generate.
For example, a bushy tree requires join operators that can join either base relations, \emph{i.e.}, triple patterns,
or intermediate relations, \emph{i.e.}, another joins.
If no available preemptable operators can meet these requirements, then the query optimizer
cannot generate bushy trees.

\section{\sagee Query Optimizer and Query Engine}\label{sec:engine}

\begin{figure}[t]
  \centering
  \input{figures/q1_optim.tex}
  \caption{Preemptable physical query execution plan produced by
  the \sagee query optimizer for $Q_1$, from Figure~\ref{query:q1}}
  \label{fig:q1_optim}
\end{figure}
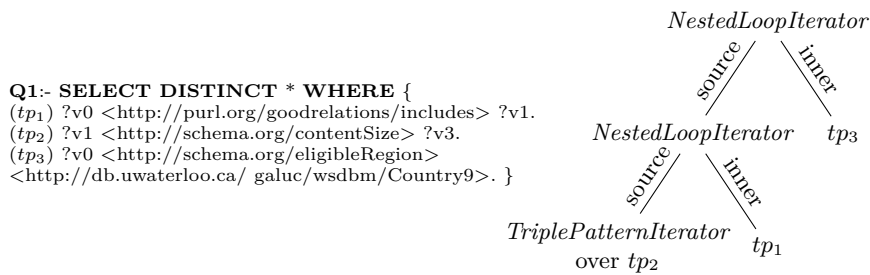

Given a BGP query, the \sagee query optimizer builds a left-linear
tree using the cardinalities of the triple patterns in the BGP.
The optimizer relies on the cardinalities of triple patterns in the query retrieved by
using indexes and the join ordering
heuristic~\cite{stocker2008sparql} for building the tree.
Figure~\ref{fig:q1_optim} shows the plan produced by the \sagee query optimizer for
query $Q_1$ of Figure~\ref{query:q1}, supposing than $|tp_2|<|tp_1|<|tp_3|$. The first triple
pattern in the plan, \emph{i.e.}, the left-most children in the plan,
is implemented using a \emph{TriplePatternIterator}, while joins are
implemented using \emph{NestedLoopIterators}.
In \sagee, both the physical query execution plan and the physical
operators must be implemented as preemptive.




\subsection{Implementation of preemptable physical query execution plan}\label{sec:sage_plan}

Algorithm~\ref{algo:preemptable_plan} presents the implementation of
the functions required for a preemptable physical query
execution. The \texttt{Stop} function simply calls \texttt{Stop} on
each operator in the tree and wait until all operators have been
stopped.  The \texttt{Save} function recursively saves each operator
in the tree while saving its structure. The \texttt{Load} function can
recursively inspect the save state produced to rebuild the tree of
operators without the re-executing of the query optimizer.

\input{./figures/preemptable_plan.tex}

These functions require to recursively stop, save or load respectively, each operator in the plan.
Thus, if these operators ensure the properties of Definition~\ref{def:preemptable_operator},
stopping, saving and loading a plan is done in $O(m)$, where $m$ is the number of operators
in the plan, \emph{i.e.}, the number of triple patterns in the query.
Consequently, all functions of Algorithm~\ref{algo:preemptable_plan}
are conform to the constraints of  Definition~\ref{def:preemptable_plan}
and implement a preemptable physical query execution plan.

Consider now the \sagee page from Figure~\ref{fig:sage_page}, which contains
a saved state of the plan of  Figure~\ref{fig:q1_optim}.
Notice that this saved state maintains the structure of the original plan.
The \emph{TriplePatternIterator} in charge of $tp_2$ has been preempted after
reading 224 solution mappings (offset), the $tp_1$ nested loop has been
preempted after examining 2 triples belonging to $mappings@tp_1$. The
$tp_3$ nested loop has been interrupted in the outer loop, so mapping
are null and offset is 0.


\subsection{Preemptable iterators for join processing}\label{sec:preempt_operators}

\sagee implements join operators in the query plan using \emph{iterators}~\cite{garcia2008database}.
An iterator is a group of
three functions: \texttt{Open}, \texttt{GetNext} and \texttt{Close}.  \texttt{Open} initializes the internal data
structures needed to perform the function, \texttt{GetNext} returns
the next results of the function and update the iterator internal
data structures, and \texttt{Close} ends the iteration and releases the
allocated ressources.
The \emph{TriplePatternIterator} implements the \emph{scan} operator
 and returns  solution mappings for a triple pattern. For \sagee, the \texttt{GetNext}  function of a \emph{TriplePatternIterator} is non interruptible.
 Later, we do not discuss this \emph{helper iterator}, details can be  found in~\cite{verborgh2016triple}.

Evaluation of Basic Graph Patterns using iterators has
already been studied in~\cite{olaf2009}. \sagee follows a similar approach for BGP evaluation
and uses \emph{NestedLoopIterators} to implements preemptable join operators.
These iterators follow the Nested Loop Join
algorithm~\cite{garcia2008database} for join processing. Given a BGP query
$B$ and the associated preemptable physical query execution plan,
a \emph{pipeline} of iterators is built, where each iterator is responsible for the evaluation of a triple pattern
from $B$.
Iterators are chained together in a \emph{pull-fashion} to respect the join ordering computed by the \sagee optimizer,
such as one iterator pulls solution mappings from its predecessor to produce results.
The iterators used by \sagee are
preemptable join operators, as defined in
Definition~\ref{def:preemptable_operator}.

\input{./figures/bgp_iterator.tex}

Algorithm~\ref{algo:bgp_iterator} gives the implementation of the
\emph{NestedLoopIterator} used by \sagee.  To produce solutions, each
iterator $I_i$ in the pipeline executes the same steps, repeatedly
until all solutions are produced:
\begin{inparaenum}[(1)]
  \item It pulls solutions mappings $\mu_c$ from its predecessor $I_{i-1}$.
  \item It applies $\mu_c$ to $tp_i$ to generate a \emph{bound pattern} $b = \mu_c\llbracket tp_i \rrbracket$.
  \item If $b$ has no solution mappings in $\mathcal{D}$,
  it tries to read again from its predecessor (jump back to Step 1).
  \item Otherwise, it reads triple matching $b$ in $\mathcal{D}$ and produces the associated solution mappings
  using a \emph{TriplePatternIterator}.
  \item When all triples matching $b$ have been read, it goes back to Step 1.
\end{inparaenum}

A \emph{NestedLoopIterator} supports preemption through the
\texttt{Stop}, \texttt{Save} and \texttt{Load} functions given in
Algorithm~\ref{algo:bgp_iterator}.
The \texttt{Stop} function waits for all non interruptible section to
have been executed before interrupting the iterator execution.  The
\texttt{Save} function saves the position of the iterator while
scanning its current bound pattern, and the \texttt{Load} function
uses these informations to resume evaluation of the bound pattern.
According to Definition~\ref{def:preemptable_operator}, the saved
state of the iterator has a size in $O(1)$.  Additionally, all
functions are in $O(1)$.  All interruptible sections
(lines~\ref{algo:line:start_synchro}-\ref{algo:line:pull_boundpattern})
that \texttt{Stop} waits for completion do not depend on the inputs,
and the state of a \emph{NestedLoopIterator} can be reloaded in
constant time if the offset function (line~\ref{algo:line:offset})
can be applied in constant time, like in
HDT~\cite{fernandez2013binary}.

Consider again the saved plan from Figure~\ref{fig:sage_page}.
The \emph{TriplePatternIterator} for $tp_2$ has been
interrupted after reading 224 solution mappings.
The \emph{NestedLoopIterator} for $tp_1$
has been interrupted after two scan in the inner loop (\emph{offset} = $2$) with $\mu_c$
bounded to \texttt{\{?v1: wm:Product11972, ?v3:"4356"\}}.
Finally, the iterator for $tp_3$
has been interrupted in the outer loop
(lines~\ref{algo:line:start_outer}-\ref{algo:line:outer_end}), \emph{i.e.},
when its pulling a solution mappings from its predecessor, so its
$\mu_c$ is not yet binded to a value and the offset is meaningless.

\section{Experimental study}\label{sec:exp_study}


We implemented the \sagee client in NodeJS and the \sagee server as a
Python web service, using HDT v1.3.2 as backend. The code, the
experimental setup and the online demo are available at the companion web
site~\footnote{\url{https://github.com/Callidon/sage-bgp}}. We run
\sagee with a quota of 75ms and a quota of 150ms to check assumptions
on the best quota values detailed in section~\ref{sec:rec}. We name
these configurations \sage-75 and \sage-150.
We compare \sagee with the following approaches:
\begin{asparadesc}
\item[Virtuoso:] Many public SPARQL endpoints rely on Virtuoso.  We
  run Virtuoso 7.2.4 with no restrictions ($VNQ$), with the quotas of
  the public DBpedia SPARQL
  endpoint~\footnote{\url{http://wiki.dbpedia.org/public-sparql-endpoint}}
  ($VQ$) and finally with quotas and pagination ($VQP$). The maximum
  query execution time is set to $120s$ and the maximum of number of
  results per query is set to $2000$. The client-side pagination
  retrieves results per page of 2000, using the
  \emph{Limit/Offset/OrderBy} technique~\cite{axel2014}.
\item[TPF:] Many data providers publish their data through TPF servers
  as Wardrobe~\footnote{\url{http://lodlaundromat.org/wardrobe/}}. We
  run the version 2.0.5 of TPF client and the version 2.2.3 of the TPF
  server.
\item[BrTPF:] In~\cite{olaf2017}, BrTPF exhibits better performances
  than other LDF interfaces. For a fair comparison with TPF, we
  re-implemented the BrTPF client~\cite{hartig2016bindings} with the
  version 2.0.5 of the TPF client.
\end{asparadesc}

\subsection{Experimental setup}

%
\textbf{Dataset and Queries:} The WatDiv
benchmark~\cite{alucc2014diversified} is designed to generate
diversified BGP queries for stress testing RDF data management
systems.  We reused the setup of the BrTPF
experiments~\cite{hartig2016bindings} based on WatDiv. The dataset
contains $10^7$
triples~\footnote{\url{http://dsg.uwaterloo.ca/watdiv/}}encoded in the
HDT format~\cite{fernandez2013binary}. The workload contains 145
SPARQL conjunctive queries with STAR, PATH and SNOWFLAKE shapes. These
queries vary in complexity, with very high and very low
selectivity. $20\%$ of queries requires more than 1s to be executed
using the virtuoso server. $7\%$ of queries produces more than 2000
results.

\textbf{Servers configurations:} We run all the servers on a machine
  with Intel(R) Xeon(R) CPU E7-8870@2.10GHz and 1.5TB RAM.  The
  clients access to the server through an HTTP proxy to ensure that
  client-server latency is kept around 50ms.  We configured a WEB
  cache NGINX of 500Mo for TPF and BrTPF, which represents
  approximately $2/3$ of the
  size of the dataset. As they don\'t use it, \sagee and Virtuoso has
  no WEB cache. We run the servers with one worker to highlight
  starvation issues. For a fair comparison, we also run \sagee and
  Virtuoso with 4 workers to study the impact of multitasking on
  starvation.

\textbf{Setup for load generation:} In order to generate load over
  servers, we rely on $n$ clients. the first $n-1$ clients, the
  loaders, continuously evaluate the $20\%$ of complex SPARQL queries of
  the workload. The last client, the measurement client, evaluates the
  145 queries of our workload. All reported results are computed on
  this last client. Except the workload, the loaders and the
  measurement client share the same configuration.

\textbf{Evaluation Metrics:} Presented results correspond to the average
  obtained of three successive execution of the queries workload.
  \begin{inparaenum}[(i)]
  \item \textit{Query timeout}: percentage of queries of the workload
    that terminate before producing complete results. The maximum query time is set to
    $120s$ as a client timeout for \sage, TPF and BrTPF and as a
    server timeout for Virtuoso.
  \item \textit{Query throughput}: the number of executed query per
    hour including timeout queries, \emph{i.e.},  145 queries of the workload
    divided by the total execution time of the workload.
  \item \textit{Preemption overhead}: is the total time taken by the
    server for stoping, saving and reloading a preemptable physical
    query execution plan.
  \item \textit{Number of HTTP requests}: is the total number of HTTP
    requests sent by a client to a server in order to evaluate a
    query.
  \item \textit{Completeness}: is the ratio between the number of
    query answers produced during the experiment and the number of complete
    results. We used Virtuoso to compute complete results.
  \end{inparaenum}

\begin{figure}[t]
  \begin{minipage}{0.5\textwidth}
    \centering
    \captionsetup{width=0.95\linewidth}
    \includegraphics[width=\textwidth]{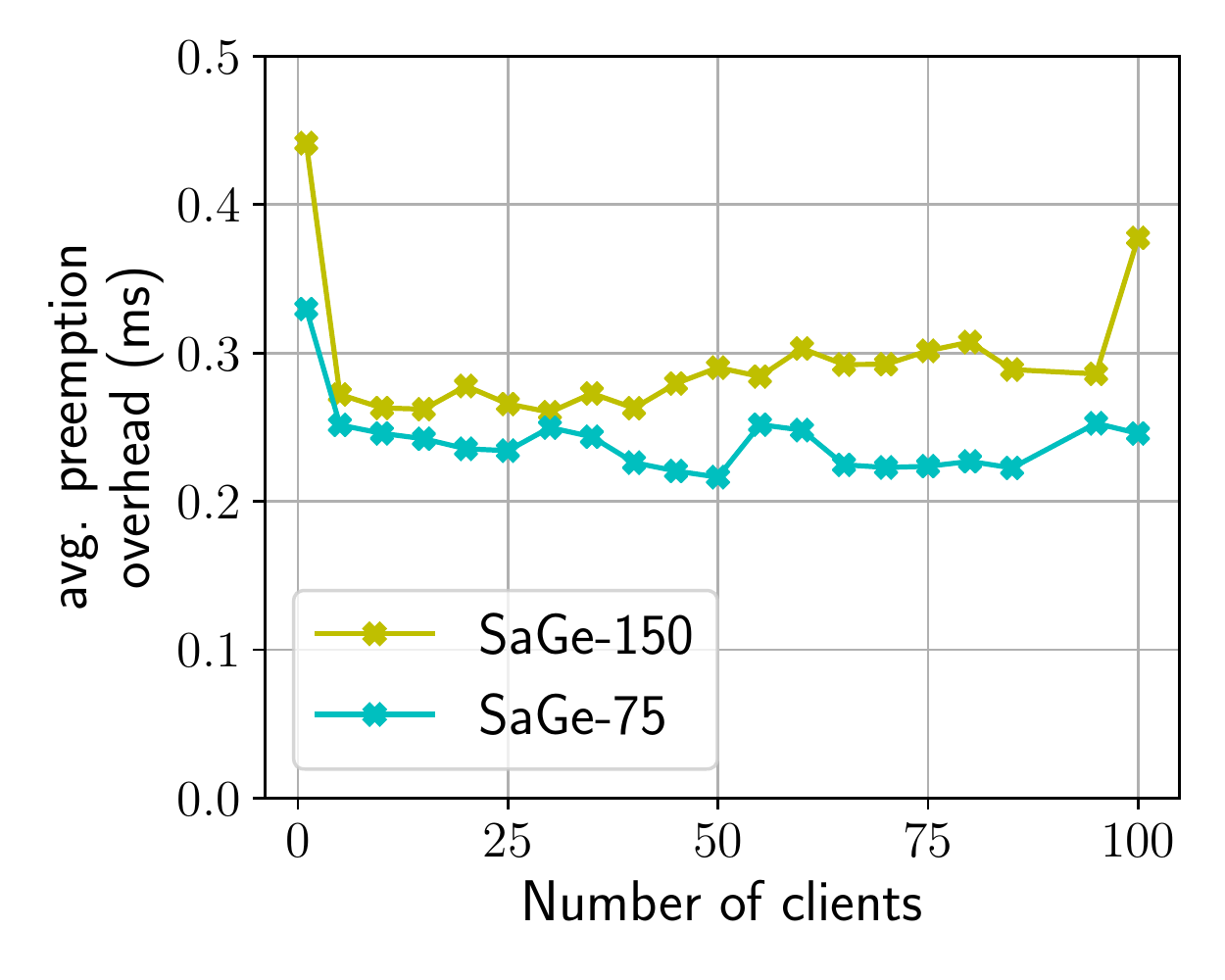}
    \caption{\sagee Preemption overhead }
    \label{fig:overhead}
  \end{minipage}\hfill
  \begin{minipage}{0.5\textwidth}
    \centering
    \captionsetup{width=0.95\linewidth}
    \includegraphics[width=\textwidth]{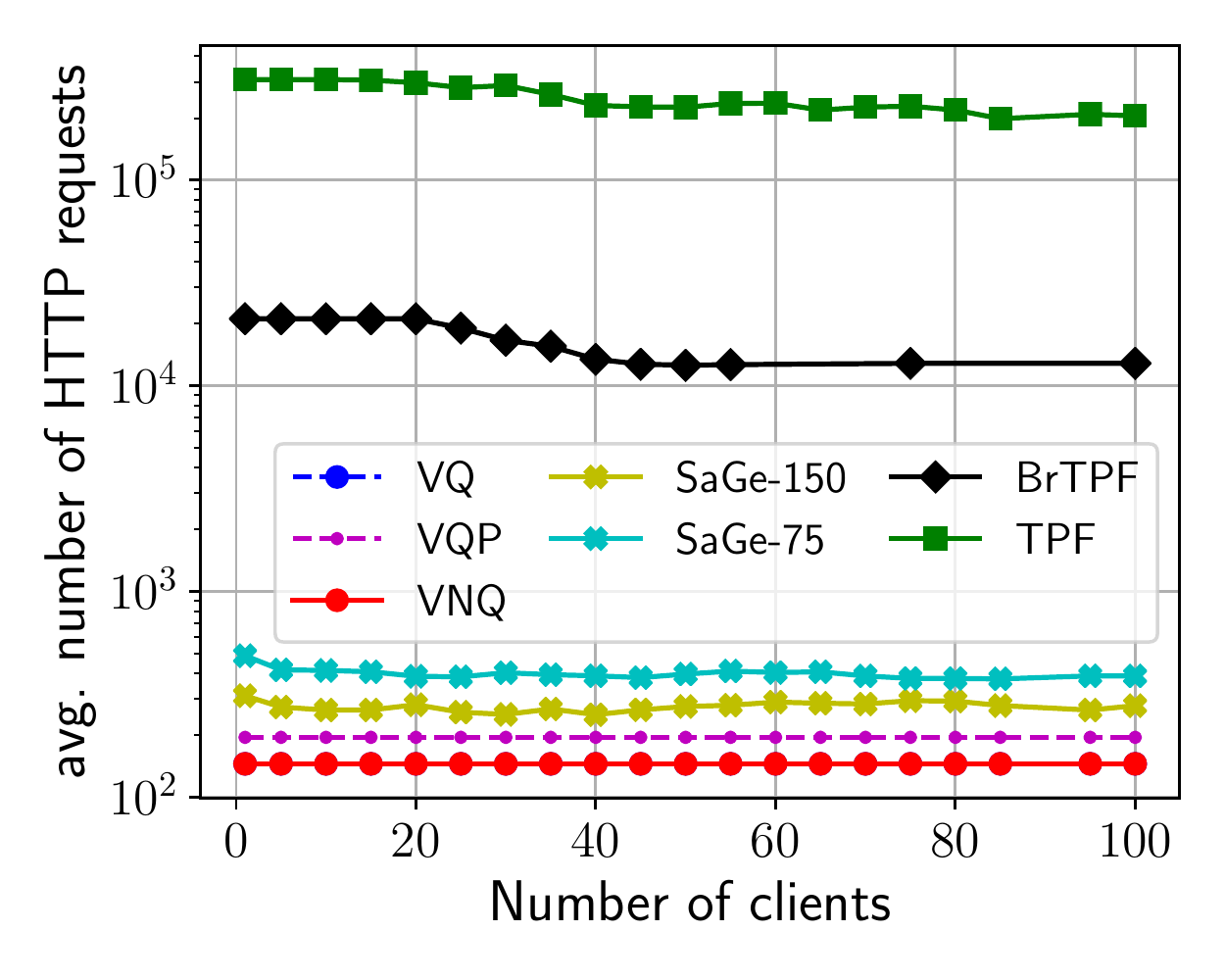}
    \caption{Average number of HTTP requests}
    \label{fig:calls}
  \end{minipage}
\end{figure}

\subsection{Experimental results}

\begin{figure}[t]
  \centering
  \begin{minipage}{0.5\textwidth}
    \centering
    \captionsetup{width=0.95\linewidth}
    \includegraphics[width=\textwidth]{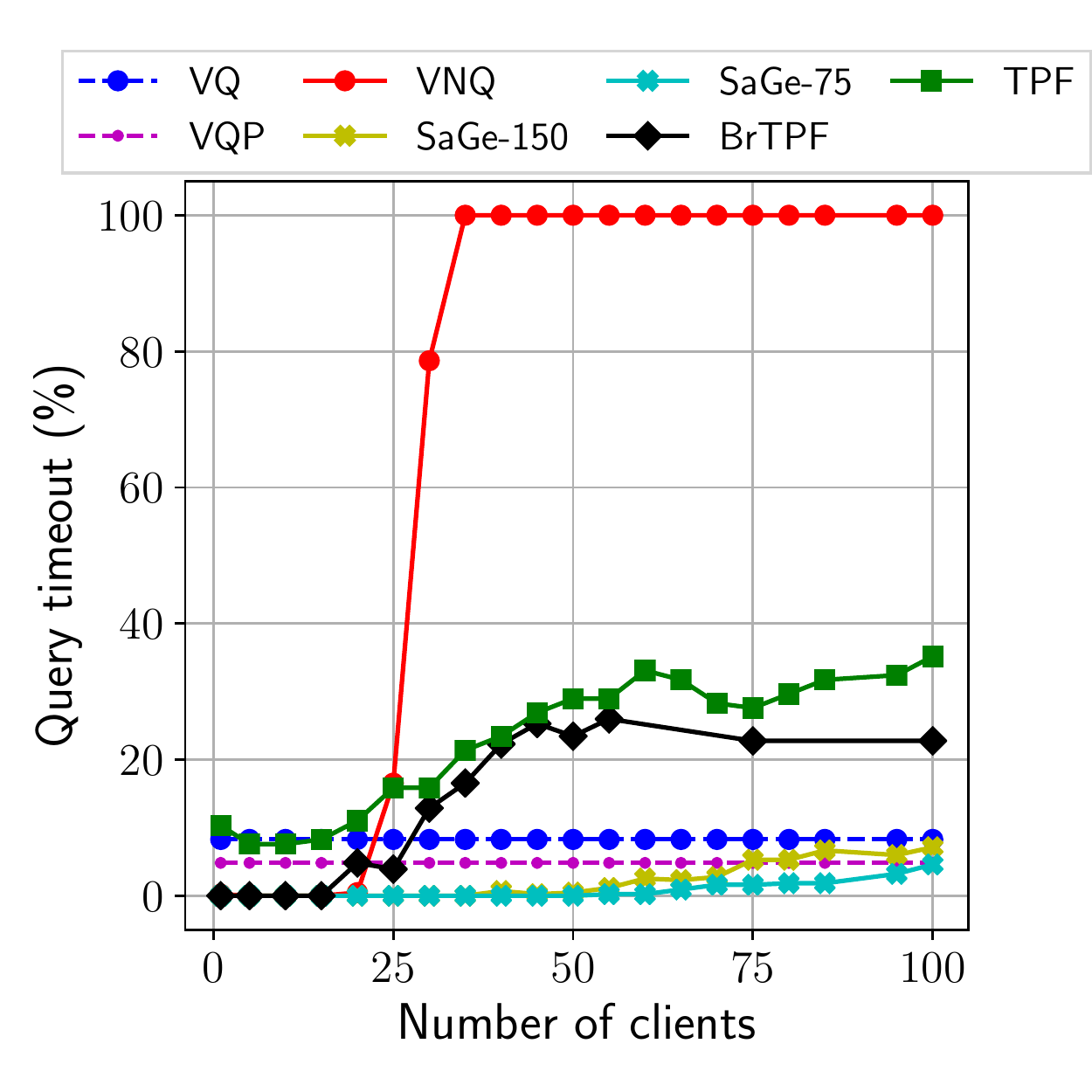}
    \caption{Average query timeout}
    \label{fig:timeouts}
  \end{minipage}\hfill
  \begin{minipage}{0.5\textwidth}
    \centering
    \captionsetup{width=0.95\linewidth}
    \includegraphics[width=\textwidth]{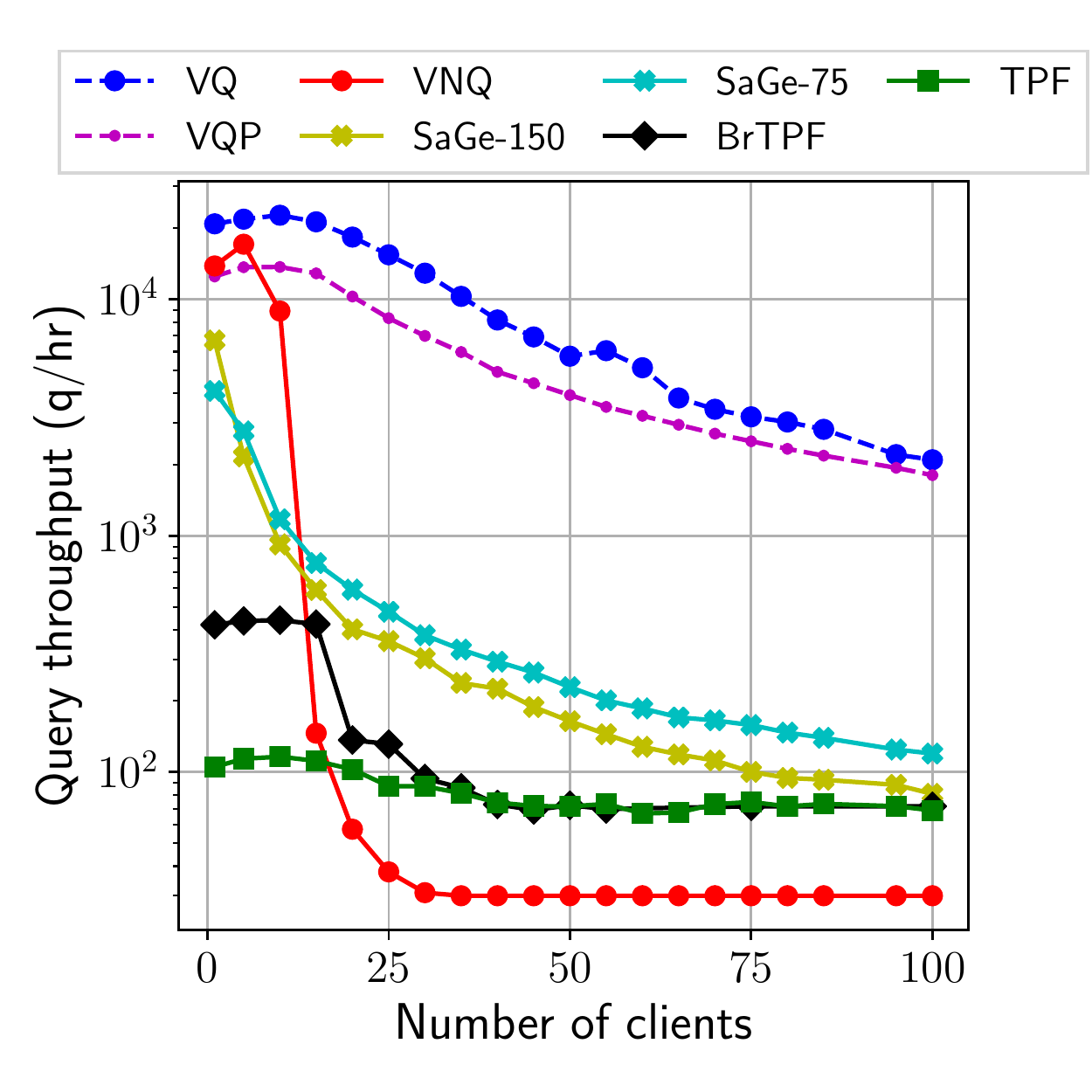}
    \caption{Average query throughput}
    \label{fig:throughput}
  \end{minipage}
\end{figure}

\begin{figure}[t]
  \centering
  \begin{minipage}{0.5\textwidth}
    \centering
    \captionsetup{width=0.95\linewidth}
    \includegraphics[width=\textwidth]{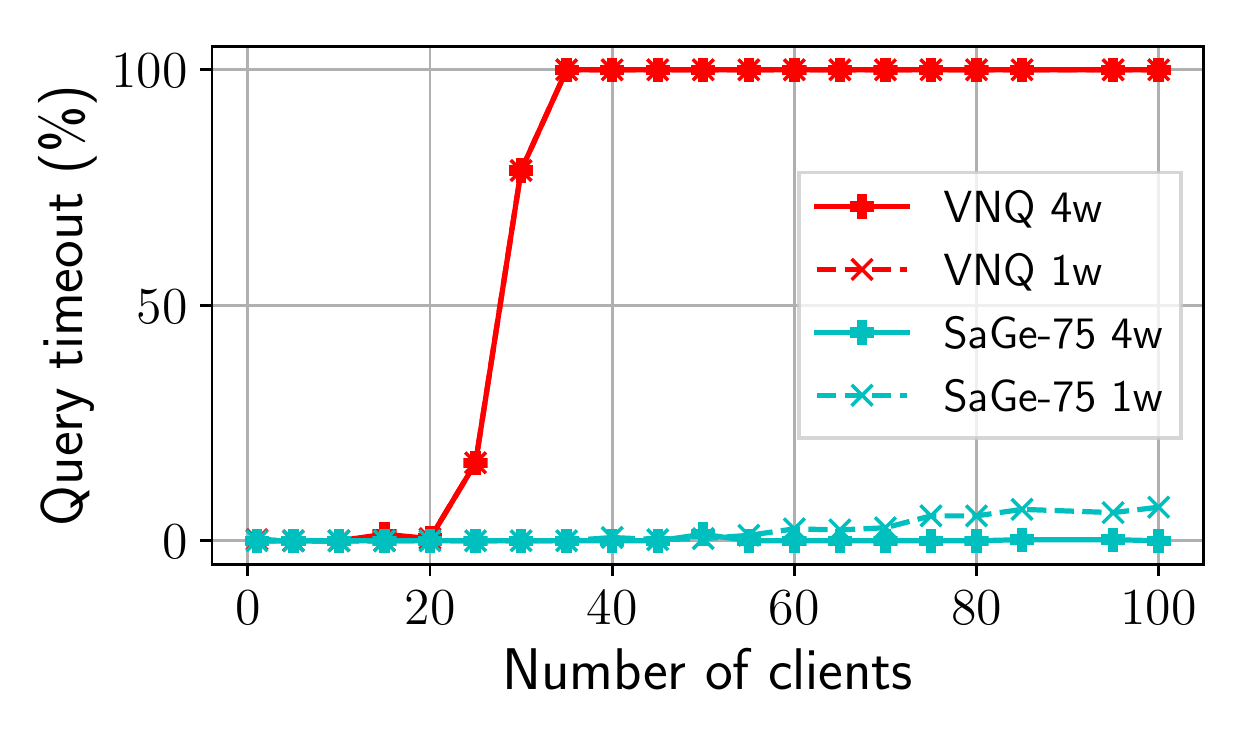}
    \caption{Average query timeout for 1 and 4 workers}
    \label{fig:timeouts4}
  \end{minipage}\hfill
  \begin{minipage}{0.5\textwidth}
    \centering
    \captionsetup{width=0.95\linewidth}
    \includegraphics[width=\textwidth]{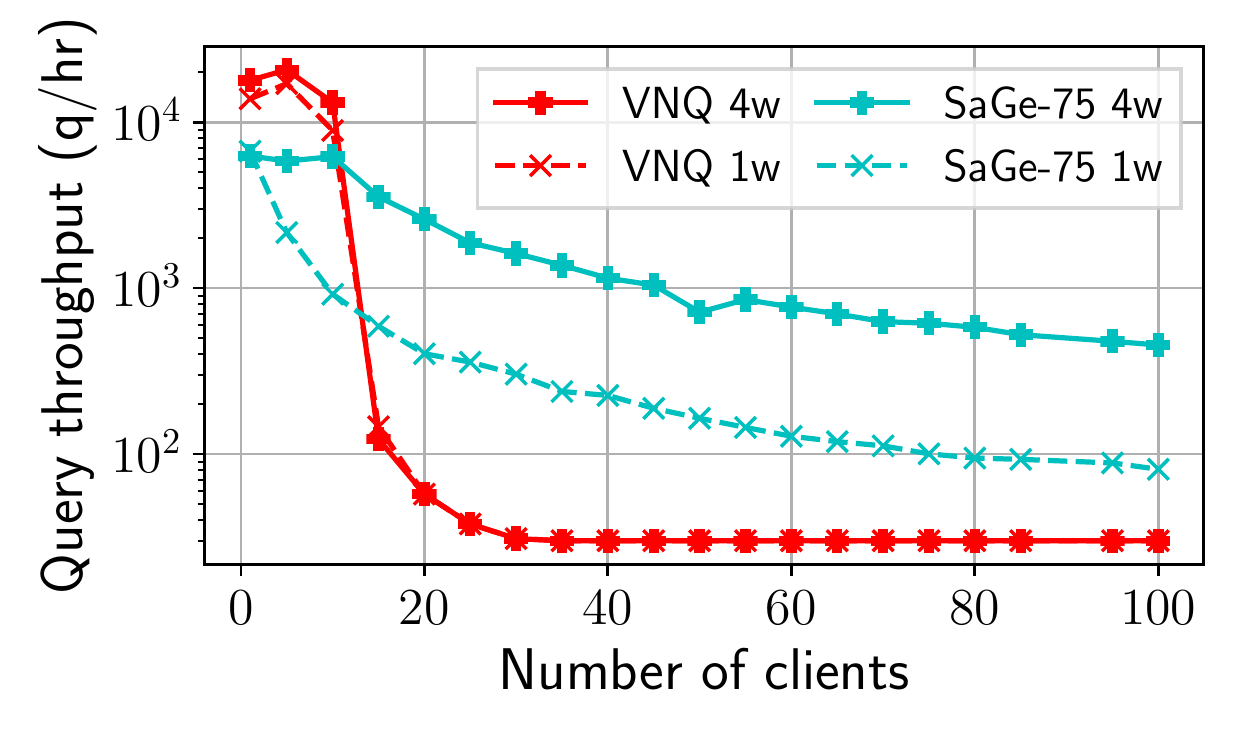}
    \caption{Average query throughput for 1 and 4 workers}
    \label{fig:throughput4}
  \end{minipage}
\end{figure}

\noindent
\textbf{Impact of the quota on \sagee}
Figure~\ref{fig:overhead} shows the average preemption overhead
obtained after evaluation of our workload by \sage-75 and \sage-150, with an
increasing load.
First, we observe that the preemption overhead do not increase with
the load in both cases. This is consistent with the properties of
\sagee preemptable physical query execution plan, as the preemption
overhead only depends on the number of triple pattern in each query.
The difference between \sage-75 and \sage-150 is in the margin of error
of measurement (<0,1ms) and are meaningless.
Second, the overhead does not exceeded $0.4\%$ of the time quota for
\sage-75 and $0.2\%$ for \sage-150, and is negligible
compared to quotas, as expected in Section~\ref{sec:rec}.
In the Figure~\ref{fig:throughput}, we compare the performances of
\sage-75 and \sage-150 in terms of timeout ratio and query throughput.
\sage-75 respect the rule of 80/20\%, explained in Section~\ref{sec:rec}, while
\sage-150 slices the workload in 84/16\%. We observe that \sage-75 has a
much better tradeoff than \sage-150 in term of query throughput and ratio of
timeout. Thus, we focus on \sage-75 for the rest of the experiments.

\noindent
\textbf{Performance analysis}
Figure~\ref{fig:timeouts} shows the ratio of query timeouts obtained
during query execution with different load configurations, up to 100
clients. Figure~\ref{fig:throughput} shows the corresponding average
throughput.

Concerning the ratio of query timeout, \sagee outperforms all
other approaches. The poor performances in throughput and timeouts of TPF and BrTPF are due to
the high data transfers. The Figure~\ref{fig:calls} confirms this observation,
\emph{i.e.}, the number of calls of \sagee is clearly lower than BrTPF and TPF.
$VQ$ exhibits constant timeout ratio, even when the server is not
loaded. Indeed, the queries that return more than 2000 results are
timed-out by the server. As only simple queries are completely
 executed, the throughput is high, but is not significative.
This is confirmed by the completeness of results: $VQ$
only delivers $20\%$ of results for the workload
whatever the number of clients. \sagee delivers complete
results up to 60 clients, which drops to $98\%$ at 100 clients.
$VQP$ also exhibits constant timeout ratio, even when not loaded.
Here, Virtuoso timeout queries because they exceeded the maximum number of row
that can be sorted by an ORDER BY clause. Consequently, complex queries in the workload cannot be
executed just relying on the SPARQL interface and, as for
$VQ$, the query throughput of $VQP$ is not significative.
$VNQ$ timeouts grow quickly up to 100\% after a load of 15 clients.
The queue of server is clearly congested with the complex queries of
the loaders. This demonstrates a poor management of preemption with
queries in the server queue.

Concerning the query throughput, $VQ$ and $VQP$ are not significative
due to the timeouts. \sagee outperforms BrTPF and TPF. $VNQ$
outperforms \sagee on the range 1 to 15 clients, However, after 15
clients, \sagee clearly outperforms $VNQ$. We conjecture that $VNQ$
produces more efficient plans with more efficient operators than
\sagee. After 15 clients, due to the congestion of the server, the
$VQ$ throughput collapses.
We rerun the same experiment with 4 workers for \sagee and $VNQ$ to
observe how more multitasking impact the results. As we see in the
Figures~\ref{fig:timeouts4} and \ref{fig:throughput4}, the general
behavior of $VNQ$ remains nearly the same, the four workers support
just a slightly more load before congestion. We observe that \sagee
performances are significantly improved with four workers; all queries
produce complete answers and the throughput is multiplied by 5.

According to these results, the \sagee approach seems to be the best
option for a public endpoint.  Indeed, only $VNQ$ delivers a better
throughput for a slightly loaded server, but a public SPARQL endpoint
without quotas is not a viable option for a data provider.


\section{Conclusion and Future Works}\label{sec:conclusion}

In this paper, we proposed \sage: a stateless preemptable SPARQL query
engine for public endpoints. Thanks to preemptable query plans and
time-sharing scheduling, \sagee tackles the problem of RDF data
availability for complex queries in public endpoints. Consequently,
\sagee provides a convenient alternative to the current practice of
copying RDF data dumps. Experimental study demonstrates that \sagee
outperforms BrTPF, TPF and Virtuoso in terms of the ratio of query
timeout.

\sagee opens several perspectives. First, in this paper, we focused on
the evaluation of conjunctive SPARQL queries.  We plan to extend
\sagee to support full SPARQL queries.
Second, we implemented the preemptable plans as simple as possible. We
think that there is room for building more efficient preemptable plans
with better preemptable operators.
Third, we used a Round-Robin scheduling strategy for its simplicity,
we plan to explore  if a more elaborated scheduling strategy~\cite{silberschatz2014operating} can
increase the performances.
Fourth, we determined the time quota statically by analyzing the
workload of the experiment. We plan to compute the quota dynamically
on server side.
Finally, we plan to extend \sagee to support federated SPARQL query
processing to overcome problems highlighted in~\cite{axel2014}.

\bibliographystyle{splncs03}
\bibliography{paper}
\end{document}

%% file: figures/sage_example.tex
 \begin{tikzpicture}[
  scale=1, transform shape,start chain=going right,node distance=0pt,
  block/.style = {rectangle, draw, text centered, minimum height=2em, text width=2cm},
  array/.style = {node distance = 0pt, draw,rectangle,minimum height=5mm, outer sep=0pt},
  round/.style = {draw, circle, minimum size=1em},
  arrow/.style = {draw,thick,-latex}]

  \node[draw,rectangle,on chain,draw=white,minimum size=1.3cm] (queue_edge) {};
  \node[rectangle split, rectangle split parts=6,
  draw, rectangle split horizontal,text height=0.7cm,text depth=0.5cm,on chain,inner ysep=0pt] (queue) {
    $Q_6$
    \nodepart{second} $Q_5$
    \nodepart{third} \dots
    \nodepart{fourth} $Q_3^{pre}$
    \nodepart{five} $Q_2^{pre}$
    \nodepart{six} $Q_1$
  };
  \fill[white] ([xshift=-\pgflinewidth,yshift=-\pgflinewidth]queue.north west) rectangle ([xshift=-20pt,yshift=\pgflinewidth]queue.south);

  \node[on chain,left=5mm of queue_edge] (dots_clients) {\vdots};
  \node[round, on chain,above=of dots_clients] (client1) {$C_1$};
  \node[round, on chain,below=1mm of dots_clients] (client2) {$C_n$};
  \draw[arrow] (client1) edge node [below,sloped,align=center] {query\\ \emph{or} next} (queue);
  \draw[arrow] (client2) edge (queue);

  \node at (queue.east) (A){};
  \draw [arrow] (A) --+(25:1.5) coordinate (B1);
  \draw [arrow] (A) --+(-25:1.5) coordinate (B3);
   \node [round, on chain] at (B1) (se1) {$w_1$};
   \node [round, on chain] at (B3) (se2) {$w_k$};
   \node [above] (dots) at (se2.north) {\vdots};


   \node[align=center,below] at (queue.south) (queue_label) {BGP Query queue};
   \node[align=center,below] at (se2.south) (server_label) {Server
     workers \\ with quotas};

   \node[draw,dashed, minimum width=1cm,
     fit=(client1) (dots_clients) (client2)] (client_box) {};
   \node [below=2mm of client_box] {\sagee Clients};

   \node[draw,dashed, minimum width=1cm,
     fit=(queue) (se1) (dots) (se2) (queue_label) (server_label)] (server_box) {};
   \node [below=2mm of server_box] {\sagee server};

   \draw[arrow](se1) |- +(0,1)  -|  (client1);
   \node [above left=1.3cm of se1] {$\langle$results, saved plan$\rangle$ (see Figure~\ref{fig:q1_page})};


\end{tikzpicture}

%% file: figures/q1_page.tex
\begin{tikzpicture}
  \tikzstyle{level 1}=[level distance=0.5cm,sibling distance=6.5cm]
  \tikzstyle{level 2}=[level distance=1.9cm, sibling distance=3.2cm]
  \node [draw] (root) {A \sagee page}
  child { node (bindings) {mappings}
  }
  child {
    node {\emph{NestedLoopIterator}}
    child {
      node {\emph{NestedLoopIterator}}
      child { node [align=center]{
        \emph{TriplePatternIterator} \\
        $tp = tp_2$ \\
        \emph{offset} = 224}
        edge from parent node [midway,sloped, above] {source}
            }
            child { node [align=center]{
              $tp = tp_1$ \\
              $\mu_c = \{$ {\scriptsize ?v1: wm:Product11972,}\\ {\scriptsize ?v3:"4356"} $\}$ \\
              \emph{offset} = 2}
              edge from parent node [midway,sloped, above] {inner}
            }
            edge from parent node [midway,sloped, above] {source}
          }
          child { node [align=center]{$tp = tp_3$ \\ $\mu_c = nil$ \\ \emph{offset} = 0  }
            edge from parent node [midway,sloped, above] {inner}
          }
          edge from parent node [midway,sloped, above] {next}
};

  \node [draw, align=center, below=2mm of bindings] (b1) {
      ?v0: wm:Offer79386, \\
      ?v1: wm:Product10014, \\
      ?v3: "5960"
  };

  \node [draw, align=center, below=2mm of b1] (b2) {
      ?v0: wm:Offer81506, \\
      ?v1: wm:Product10066, \\
      ?v3: "6563"
  };

  \path [draw]
    (bindings) edge (b1)
    (b1) edge (b2);

  \node [align=center, below=2mm of b2] (bdots) {$\vdots$};


  \end{tikzpicture}

%% file: figures/q1_optim.tex

\begin{tikzpicture}
 \tikzstyle{level 1}=[sibling distance=2cm]
  \tikzstyle{level 2}=[sibling distance=2cm]
  \node (root ){\emph{NestedLoopIterator}}
  child {
    node (nlj) {\emph{NestedLoopIterator}}
    child { node [align=center]{\emph{TriplePatternIterator}\\over $tp_2$}
      edge from parent node [midway,sloped, above] {source}
    }
    child { node [align=center]{$tp_1$}
      edge from parent node [midway,sloped, above] {inner}
    }
    edge from parent node [midway,sloped, above] {source}
  }
  child { node [align=center]{$tp_3$}
    edge from parent node [midway,sloped, above] {inner}
  };

 \node [left = 0.5cm of nlj, align=left,font=\scriptsize] {
 \textbf{Q1}:- \textbf{SELECT DISTINCT} * \textbf{WHERE} \{ \\
  $(tp_1)$ ?v0 <http://purl.org/goodrelations/includes> ?v1.\\
  $(tp_2)$ ?v1 <http://schema.org/contentSize> ?v3. \\
  $(tp_3)$ ?v0 <http://schema.org/eligibleRegion> \\
    <http://db.uwaterloo.ca/~galuc/wsdbm/Country9>. \}
 };

  \end{tikzpicture}

%% file: figures/preemptable_plan.tex
\begin{algorithm}
  \SetAlgoVlined
  \SetKwInput{Input}{Require}
  \SetKwBlock{Synchro}{synchronized}{{end synchronized}}
  \SetKwFor{Foreach}{for each}{do}{endfor}
  \SetKwProg{Fn}{Function}{\string:}{}
  \SetKwProg{Proc}{Procedure}{\string:}{}
  \SetKwProg{Event}{Event}{\string:}{}
  \SetKwComment{tcp}{}{}%
  \SetKwComment{tcc}{// }{}%

  \DontPrintSemicolon
  \SetInd{0.55em}{0.55em}
  \caption{\sagee preemptable physical query execution plan}
  \label{algo:preemptable_plan}

  \Input{
    $\mathcal{P}$: tree of operators,
    $S$: saved plan (as generated by \emph{Plan.Save})
  }

  \setlength\columnsep{-95pt}
  \begin{multicols}{2}

  \Fn{Plan.Stop($\mathcal{P}$)}{
    \textbf{let} $op \leftarrow \mathcal{P}$ \;
    \While{$op \neq nil$}{
      \textbf{Call} $op.Stop()$\;
      $op \leftarrow op.predecessor$ \;
    }
  }
  \BlankLine
  \Fn{Plan.Save($\mathcal{P}$)}{
    \If{$\mathcal{P} = nil$}{
      \Return $nil$ \;
    }
    \textbf{let} $s \leftarrow \mathcal{P}.Save()$ \;
    \textbf{let} $pred \leftarrow Plan.Save(\mathcal{P}.predecessor)$ \;
    \Return $\langle pred, s \rangle$ \;
  }

  \BlankLine
  \vfill
  \columnbreak

  \Fn{Plan.Load($S$)}{
    \textbf{let} $\langle pred, s \rangle \leftarrow S$ \;
    \textbf{let} $\langle tp, \mu, n \rangle \leftarrow s$\;
    \eIf{$pred = nil$}{
      $op \leftarrow TriplePatternIterator$ over $tp$ in $\mathcal{D}$ \;
    }{
      $I_{pred} \leftarrow Plan.Load(pred)$ \;
      $op \leftarrow NestedLoopIterator(tp, \mathcal{D}, I_{pred})$ \;
    }
    $op.Load(tp, \mu, n)$ \;
    \Return $op$
  }
  \end{multicols}
  \BlankLine
  \BlankLine
\end{algorithm}

%% file: figures/bgp_iterator.tex
\begin{algorithm}[t]
  \SetAlgoVlined
  \SetKwInput{Data}{Require}
  \SetKwBlock{Synchro}{non interruptible}{{end synchronized}}
  \SetKwFor{Foreach}{for each}{do}{endfor}
  \SetKwProg{Fn}{Function}{\string:}{}
  \SetKwProg{Proc}{Procedure}{\string:}{}
  \SetKwProg{Event}{Event}{\string:}{}
  \SetKwComment{tcp}{}{}%
  \SetKwComment{tcc}{// }{}%

  \DontPrintSemicolon
  \SetInd{0.55em}{0.55em}
  \caption{A \textit{NestedLoopIterator} $I_i$: a preemptable join operator used by \sagee}
  \label{algo:bgp_iterator}

  \Data{
    $tp_i$: triple pattern evaluated by $I_i$,
    $\mathcal{D}$: RDF dataset queried,
    $I_{i-1}$: iterator responsible for the evaluation of $tp_{i-1}$.
  }

  \setlength\columnsep{0pt}
  \begin{multicols*}{2}

  \Fn{Open()}{
    $I_{i-1}.Open()$ \;
    $\mu_c \leftarrow nil$ \;
    $I_{find} \leftarrow$ nil \; 
  }

  \Fn{GetNext()}{
    \While{$I_{find}.GetNext() = nil $}{ \label{algo:line:start_outer}
      $\mu_c \leftarrow I_{i-1}.GetNext()$ \; \label{algo:line:pull_mappings}
      \If{$\mu_c =  nil $}{
        \Return $nil $ \;
      }
      $I_{find} \leftarrow$ \emph{TriplePatternIterator}\\ over
      $\mu_c\llbracket tp_i \rrbracket_\mathcal{D}$\; \label{algo:line:get_cursor} \label{algo:line:outer_end}
    }
    \Synchro{ \label{algo:line:start_synchro}
      \textbf{let} $\mu \leftarrow I_{find}.GetNext()$ \;
      \Return $\mu \medcup \mu_c$ \; \label{algo:line:pull_boundpattern}
    }
  }

  \Fn{Close()}{
    $I_{i-1}.Close()$ \;
  }

  \BlankLine
  \vfill
  \columnbreak

  \Fn{Stop()}{
    \textbf{Wait} until all \textbf{non interruptible} sections
    have been evaluated \;
    \textbf{Interrupt} ongoing $GetNext()$ calls \;
  }

  \Fn{Save()}{
    \textbf{let} $n \leftarrow$ the number of triples already read by $I_{find}$ \;
    \Return $\langle tp_i, \mu_c, n \rangle$ \;
  }

  \Fn{Load($tp', \mu', n$)}{
    $tp_i \leftarrow tp'$ \;
    \If{$\mu' \neq nil $}{
      $\mu_c \leftarrow \mu'$ \;
      $I_{find} \leftarrow$ \emph{TriplePatternIterator} over\\
      $\mu_c\llbracket tp_i \rrbracket_\mathcal{D}$\;
      \If{$n > 0$}{
        \textbf{Skip} the $n$ first results of $I_{find}$\; \label{algo:line:offset}
      }
    }
  }
\end{multicols*}
  \BlankLine
  \BlankLine
\end{algorithm}

%% file: paper.bbl
\begin{thebibliography}{10}
\providecommand{\url}[1]{\texttt{#1}}
\providecommand{\urlprefix}{URL }

\bibitem{alucc2014diversified}
Alu{\c{c}}, G., Hartig, O., {\"O}zsu, M.T., Daudjee, K.: Diversified stress
  testing of rdf data management systems. In: International Semantic Web
  Conference. pp. 197--212. Springer (2014)

\bibitem{axel2014}
Aranda, C.B., Polleres, A., Umbrich, J.: Strategies for executing federated
  queries in {SPARQL1.1}. In: The Semantic Web - {ISWC} 2014 - 13th
  International Semantic Web Conference. pp. 390--405 (2014)

\bibitem{bizer2009linked}
Bizer, C., Heath, T., Berners{-}Lee, T.: Linked data - the story so far. Int.
  J. Semantic Web Inf. Syst.  5(3),  1--22 (2009)

\bibitem{brockmeyer1948life}
Brockmeyer, E., Halstrm, H., Jensen, A., Erlang, A.K.: The life and works of ak
  erlang.  (1948)

\bibitem{buil2013sparql}
Buil-Aranda, C., Hogan, A., Umbrich, J., Vandenbussche, P.Y.: {SPARQL}
  web-querying infrastructure: Ready for action? In: International Semantic Web
  Conference. pp. 277--293. Springer (2013)

\bibitem{fernandez2013binary}
Fern{\'a}ndez, J.D., Mart{\'\i}nez-Prieto, M.A., Guti{\'e}rrez, C., Polleres,
  A., Arias, M.: Binary {RDF} representation for publication and exchange
  ({HDT}). Web Semantics: Science, Services and Agents on the World Wide Web
  19,  22--41 (2013)

\bibitem{garcia2008database}
Garcia-Molina, H., Ullman, J.D., Widom, J.: Database systems: the complete
  book. Pearson Education India (2008)

\bibitem{olaf2009}
Hartig, O., Bizer, C., Freytag, J.C.: Executing {SPARQL} queries over the web
  of linked data. In: The Semantic Web - {ISWC} 2009, 8th International
  Semantic Web Conference. pp. 293--309 (2009)

\bibitem{hartig2016bindings}
Hartig, O., Buil-Aranda, C.: Bindings-restricted triple pattern fragments. In:
  Proceedings of the 15th International Conference on Ontologies, Databases,
  and Applications of Semantics (ODBASE) (2016)

\bibitem{olaf2017}
Hartig, O., Letter, I., P{\'{e}}rez, J.: A formal framework for comparing
  linked data fragments. In: The Semantic Web - {ISWC} 2017 - 16th
  International Semantic Web Conference. pp. 364--382 (2017)

\bibitem{minierESWC2018ulysses}
Minier, T., Skaf{-}Molli, H., Molli, P., Vidal, M.: Intelligent clients for
  replicated triple pattern fragments. In: Proceedings of the 15th Extended
  Semantic Web Conference ({ESWC}) (2018)

\bibitem{PangCL93}
Pang, H., Carey, M.J., Livny, M.: Partially preemptive hash joins. In:
  Proceedings of the 1993 {ACM} {SIGMOD} International Conference on Management
  of Data, Washington, D.C., May 26-28, 1993. pp. 59--68 (1993)

\bibitem{silberschatz2014operating}
Silberschatz, A., Galvin, P.B., Gagne, G.: Operating system concepts
  essentials. John Wiley \& Sons, Inc. (2014)

\bibitem{stocker2008sparql}
Stocker, M., Seaborne, A., Bernstein, A., Kiefer, C., Reynolds, D.: Sparql
  basic graph pattern optimization using selectivity estimation. In:
  Proceedings of the 17th international conference on World Wide Web. pp.
  595--604. ACM (2008)

\bibitem{Stonebraker81}
Stonebraker, M.: Operating system support for database management. Commun.
  {ACM}  24(7),  412--418 (1981)

\bibitem{van2015substring}
Van~Herwegen, J., De~Vocht, L., Verborgh, R., Mannens, E., Van~de Walle, R.:
  Substring filtering for low-cost linked data interfaces. In: International
  Semantic Web Conference. pp. 128--143. Springer (2015)

\bibitem{vander2015opportunistic}
Vander~Sande, M., Verborgh, R., Van~Herwegen, J., Mannens, E., Van~de Walle,
  R.: Opportunistic linked data querying through approximate membership
  metadata. In: International Semantic Web Conference. pp. 92--110. Springer
  (2015)

\bibitem{verborgh2016triple}
Verborgh, R., Vander~Sande, M., Hartig, O., Van~Herwegen, J., De~Vocht, L.,
  De~Meester, B., Haesendonck, G., Colpaert, P.: Triple pattern fragments: A
  low-cost knowledge graph interface for the web. Web Semantics: Science,
  Services and Agents on the World Wide Web  37,  184--206 (2016)

\bibitem{WIERMAN201139}
Wierman, A.: Fairness and scheduling in single server queues. Surveys in
  Operations Research and Management Science  16(1),  39 -- 48 (2011)

\end{thebibliography}
